\documentclass[11pt,a4paper]{article}
\usepackage[T1]{fontenc}
\usepackage[utf8]{inputenc}
\usepackage{lmodern}
\usepackage[margin=25mm]{geometry}
\usepackage{amsmath,amssymb,amsthm,mathtools}
\usepackage{microtype,graphicx,booktabs,tabularx,array,enumitem}
\usepackage{xcolor,tikz,pgfplots}
\pgfplotsset{compat=1.18}
\usetikzlibrary{arrows.meta,positioning}
\usepackage[colorlinks=true,linkcolor=blue!45!black,citecolor=blue!45!black,urlcolor=blue!45!black]{hyperref}
\hypersetup{pdftitle={Exact ballistic energy transport and emergent XXZ dynamics in an integrable three-state chain},pdflang={en}}
\setlist{nosep,leftmargin=1.8em}
\numberwithin{equation}{section}

\newcommand{\I}{\mathbb I}
\newcommand{\C}{\mathbb C}
\newcommand{\R}{\mathbb R}
\newcommand{\tr}{\operatorname{tr}}
\newcommand{\Tr}{\operatorname{Tr}}
\newcommand{\spec}{\operatorname{spec}}
\newcommand{\rank}{\operatorname{rank}}
\newcommand{\ket}[1]{|#1\rangle}
\newcommand{\bra}[1]{\langle#1|}
\title{Exact ballistic energy transport and emergent XXZ dynamics in an integrable three-state chain}
\newcommand{\ManuscriptAuthorNames}{Hanbing Liang and Fujun Liu}
\newcommand{\SharedAffiliation}{Nanophotonics and Biophotonics Key Laboratory
of Jilin Province, School of Physics, Changchun University of Science and
Technology, Changchun 130022, China}

\author{\ManuscriptAuthorNames\thanks{Corresponding author: Fujun Liu.
E-mail: \href{mailto:fjliu@cust.edu.cn}{fjliu@cust.edu.cn}.}\\[0.6em]
\parbox{0.95\textwidth}{\centering\small\SharedAffiliation}}
\date{}
\begin{document}
\maketitle
\begin{abstract}
We investigate the coupling dependence of ballistic energy transport and the emergent spin dynamics in an integrable Hermitian three-state chain that connects a clock interaction to a highly degenerate flag limit. By constructing a regular $R$-matrix to establish a globally conserved energy current, we analytically evaluate its full variance to obtain the exact, strictly positive leading high-temperature coefficient of the thermal Drude weight and the ballistic growth rate of the energy-correlation second moment. In the strong-coupling limit, the degeneracy is lifted by virtual transitions of a delocalized third-color spectator state, which generates an effective spin-$1/2$ XXZ Hamiltonian with anisotropy $\Delta = -1/2$ and fundamentally selects the all-active two-color sector as the true ground state. For periodic boundaries, this virtual spectator motion introduces a positive length-changing XXZ supercharge squared that, for $L\ge4$, strictly annihilates all states within a finite, length-independent energy interval above the ground state. Consequently, we rigorously prove that the full periodic effective theory perfectly replicates the exact low-energy XXZ spectrum, including all state multiplicities, as well as its macroscopic bulk free-energy density.
\end{abstract}


\section{Introduction}

The interplay between conservation laws and ballistic transport, alongside the emergence of effective spin dynamics in strongly coupled degenerate limits, represents a central focus in the study of integrable quantum many-body systems. For commensurate easy-plane anisotropies of the spin-$1/2$ XXZ chain, quasilocal charges provide bounds on the high-temperature spin Drude weight \cite{prosen}, whereas exactly conserved energy currents allow the thermal Drude weight to be determined directly by static current fluctuations \cite{klumper,zotos,transportreview}. Fateev-Zamolodchikov clock models \cite{fz} offer a rich three-state framework for exploring these phenomena, with previous studies employing spin-vertex mappings \cite{martins}, chiral Potts Yang-Baxter constructions \cite{zhang}, toroidal Bethe equations \cite{martinspotts}, boundary-driven matrix-product-state simulations \cite{clocktransport,clocktemperature}, and generalized hydrodynamics \cite{mazza}. In particular, the thermal Drude weight has been evaluated along an integrable chiral clock line using Mazur bounds and time-dependent density matrix renormalization group methods, although the generic energy current remains nonconserved \cite{mannadrude}. To systematically address both the transport and low-energy spectral problems, we investigate a one-parameter Hermitian family of an integrable three-state chain that smoothly connects a clock interaction at $q^2=4$ to an extensively degenerate flag limit at large $|q|$ \cite{flag}.

To resolve the transport properties across the full three-state Hilbert space, we construct a regular $R$-matrix and rigorously prove the global conservation of the total energy current at every finite real coupling $q$. By evaluating the full variance of this conserved current via local traces, we capture the crucial adjacent-current correlations that dictate the coupling dependence of the ballistic response. This exact calculation yields a strictly positive leading high-temperature thermal Drude weight throughout the entire interpolation family. At the common clock point, our analytical result exactly reproduces the numerical coefficient of Ref.~\cite{mannadrude} after applying an explicit energy rescaling. Furthermore, through the continuity equation, this same local variance strictly determines the exact quadratic time dependence of the infinite-temperature energy-correlation second moment. Consequently, a unified local calculation comprehensively fixes both the amplitude of the ballistic thermal Drude weight and the spatial spreading rate of equilibrium energy fluctuations.

When approaching the strongly coupled flag limit, identifying the low-energy degrees of freedom requires resolving the splitting of the extensively degenerate leading ground space induced by virtual transitions. We demonstrate that this degenerate manifold contains both an active two-color spin sector and delocalized states of a third color acting as a spectator. On an open chain, we derive the effective Hamiltonians and find that the zero-spectator and one-spectator sectors form spin-$1/2$ XXZ chains with an anisotropy of $\Delta=-1/2$. A comparison of these effective blocks reveals that the all-active two-color sector possesses the lowest ground energy, separated from the spectator states by a strictly positive energy cost. Closing the chain into a uniform real-gauge ring alters the available paths for virtual spectator motion, generating configuration cycles that introduce a positive square of the length-changing XXZ supercharge into the active sector \cite{hagendorf,matsui}. By utilizing an adjacent-length spectral relation and bounding the periodic resolvents, we rigorously prove that this virtual correction annihilates every XXZ state within a length-independent interval of width $(7-\sqrt{33})/4$ above the ground energy for chains of length $L \ge 4$. The full periodic effective theory thus preserves the exact XXZ low-energy spectrum, including state multiplicities, and yields a macroscopic bulk free-energy density identical to that of the XXZ model.

The remainder of this paper is organized as follows. Section~\ref{sec:model} defines the Hamiltonian and establishes its underlying integrability structure. Section~\ref{sec:transport} presents the derivation of the exact transport coefficient and the energy-correlation second moment. Sections~\ref{sec:open} and \ref{sec:periodic} detail the resolution of the open and periodic effective theories, respectively, concluding with an analysis of their physical implications. Finally, Appendices~\ref{app:spectral} through \ref{app:periodic} provide the explicit polynomial $R$-matrix constructions and contain the rigorous mathematical proofs for both the transport and effective Hamiltonian results.

\section{The three-state chain}\label{sec:model}
We first specify the local interaction and the conventions used
for both calculations. The charge-sector form makes the discrete
symmetry explicit, while a real form obtained by a local unitary
transformation will be useful for evaluating traces and identifying
the strong-coupling degrees of freedom.

\subsection{Local interaction and symmetries}
Let \(\omega=e^{2\pi i/3}\), \(q\in\R\), \(t=q^2\), and
\(a=(t-1)/3\). The on-site basis is \(\ket0,\ket1,\ket2\).
In the two-site charge sectors
\[
 \mathcal H_0=(00,12,21),\qquad
 \mathcal H_1=(01,10,22),\qquad
 \mathcal H_2=(02,11,20),
\]
define \(h_q=D_0\oplus D_1\oplus D_2\), where
\begin{align}
D_0&=\begin{pmatrix}
4a&2&2\omega^2\\2&-2a&2\omega^2\\2\omega&2\omega&-2a
\end{pmatrix},\nonumber\\
D_1&=\begin{pmatrix}
a&(t-2)\omega^2&q\\(t-2)\omega&a&q\omega\\
q&q\omega^2&-2a
\end{pmatrix},\qquad
D_2=\begin{pmatrix}
a&q\omega&(t-2)\omega\\q\omega^2&-2a&q\\
(t-2)\omega^2&q&a
\end{pmatrix}.
\label{eq:density}
\end{align}
The density is Hermitian and traceless. It commutes with \(Z\otimes Z\),
where \(Z=\operatorname{diag}(1,\omega,\omega^2)\), and is balanced:
\begin{equation}
 \Tr_1h_q=\Tr_2h_q=\operatorname{diag}(2t-2,1-t,1-t).
\end{equation}
Conjugation by \(\operatorname{diag}(1,-1,-1)^{\otimes2}\)
changes \(q\) to \(-q\). The transport coefficients and spectra
considered below therefore depend on \(t\).
The off-diagonal entries include pair conversions such as
\(00\leftrightarrow12\), so the separate color occupations are
not conserved.

\subsection{Integrability and the energy current}
The Yang--Baxter equation gives the local conservation law for
the energy current. Appendix~\ref{app:spectral} constructs a polynomial \(R\)-matrix
\(\mathcal R_q(x)\), regular at \(x=1\), satisfying
\begin{equation}
\mathcal R_{12}(x)\mathcal R_{23}(xy)\mathcal R_{12}(y)
=\mathcal R_{23}(y)\mathcal R_{12}(xy)\mathcal R_{23}(x).
\label{eq:yb}
\end{equation}
Its trace-normalized form in the additive spectral parameter is
\begin{equation}
\check R_q(u)=
\frac{9\mathcal R_q(e^{i\sqrt3 u})}
{\tr\mathcal R_q(e^{i\sqrt3 u})}
,\qquad
\check R_q(0)=\I,\qquad \check R_q'(0)=h_q
\label{eq:regular}
\end{equation}
and is analytic near \(u=0\).
Regularity identifies \(h_q\) as the local Hamiltonian density.
Differentiating the Yang--Baxter equation gives
\begin{equation}
 F(h_q):=[h_{12}+h_{23},[h_{12},h_{23}]]=Y_{23}-Y_{12}
\label{eq:resh}
\end{equation}
with the explicit two-site choice \eqref{eq:Ycertificate}.
The right-hand side is a difference of translated local operators.
It therefore cancels on a periodic chain, providing the conservation
law used in Sec.~\ref{sec:transport}.

\subsection{A real representation and two reference limits}
To display the interaction in a form suited to these calculations, set
\(S=\operatorname{diag}(1,1,\omega)\) and conjugate site \(j\) by
\(G_j=Z^jS\). On every open interval,
\begin{equation}
\widetilde h_q=v\otimes\I+\I\otimes v+\mathcal T,\qquad
v=\frac{t-1}{3}\operatorname{diag}(2,-1,-1),
\label{eq:real}
\end{equation}
where, with indices understood modulo three,
\begin{equation}
\mathcal T=\sum_{i,j=0}^2N_{ij}
(e_i\otimes e_j^\dagger+e_i^\dagger\otimes e_j),\qquad
e_i=\ket{i+1}\bra i,\qquad
N=\begin{pmatrix}t-2&q&2\\q&2&q\\2&q&t-2\end{pmatrix}.
\label{eq:N}
\end{equation}
This transformation separates the diagonal one-site term \(v\)
from the color-changing interaction \(\mathcal T\).
The helix closes on a ring when \(3\mid L\); otherwise it leaves
a closing-bond twist. Length-dependent boundary twists also occur
when bulk phases are removed in the Potts chains of
Ref.~\cite{martinspotts}.
For the transport calculation, all required traces involve
open intervals of at most four sites, and the resulting identities
hold for every \(L\ge5\).

\label{sec:referencepoints}
At \(q=2\), every entry of \(N\) equals two and
\(v=Z+Z^\dagger\). With the three-state shift
\(\mathsf X=\sum_i e_i\), the real-gauge density becomes
\begin{equation}
\widetilde h_2=(Z+Z^\dagger)\otimes\I+
\I\otimes(Z+Z^\dagger)
+2(\mathsf X\otimes\mathsf X^\dagger+
\mathsf X^\dagger\otimes\mathsf X).
\label{eq:clockpoint}
\end{equation}
Its periodic sum is the Fateev--Zamolodchikov clock interaction
in this normalization \cite{fz}. The sign-related point \(q=-2\)
is covered by the on-site unitary given above.

The eigenvalues \(0,t+2,t-4\) of \(N\) distinguish the
interaction from its clock limit through the operator Schmidt rank.
The six traceless operators \(e_i,e_i^\dagger\) are linearly
independent, so the connected interaction \(\mathcal T\) has
operator Schmidt rank \(2\rank N\): four for \(t\ne4\), and two
at the clock point. Here this rank is the minimum number of
products of single-site operators needed to express
\(\mathcal T\). The clock interactions and the extension in
Eq.~(100) of Ref.~\cite{martins} have rank at most two, since
their right-site factors span only two clock operators.
Independent changes of basis on the two sites and nonzero overall
scaling preserve this rank; scalar and one-site terms disappear
on taking the connected part. Thus the general coupling in
\eqref{eq:N} cannot be obtained from those interactions by local
or staggered basis changes, spectral rescaling, or a boundary twist.

For the other limit, write
\(\widetilde h_q=q^2h^{(2)}+qh^{(1)}+h^{(0)}\).
Let \(r=\ket1\bra1+\ket2\bra2\), let \(\mathsf P\) exchange the two
three-state spaces, and set
\(\mathsf P_{\rm act}=(r\otimes r)\mathsf P\). Then
\begin{equation}
h^{(2)}=\frac13\I+\mathsf P-\mathsf P_{\rm act}-r\otimes r.
\label{eq:flagidentification}
\end{equation}
After relabeling the two active basis states, this is model
\(\mathrm{II}_{-}\) with flag \((3,2)\), \(a_0=1/3\), and \(b_0=1\)
in Eq.~(2.15) of Ref.~\cite{flag}. Its positive local form is displayed
in \eqref{eq:flag}. The subleading terms determine the dynamics
within this degenerate space and will be treated in Sec.~\ref{sec:open}.

\section{Energy transport at finite coupling}\label{sec:transport}
We now use the local Yang--Baxter identity to determine the
high-temperature transport weight. The calculation separates into
two steps: conservation removes the time dependence of the total
current correlation, and local traces determine its amplitude.
Throughout this section \(q\) is a fixed finite real coupling.

\subsection{The conserved energy current}
For a periodic chain, define
\begin{equation}
H_L=\sum_x h_x,\qquad
j_x=i[h_{x-1},h_x],\qquad J_L=\sum_xj_x,\qquad
\langle O\rangle_0=3^{-L}\Tr O.
\label{eq:current}
\end{equation}
Here \(h_x\) acts on \(x,x+1\), and time evolution uses
\(\hbar=1\) and unit lattice spacing.
The continuity equation \(\dot h_x=j_x-j_{x+1}\) identifies \(j_x\)
as the local energy current.

For every real \(q\) and \(L\ge5\), the total current satisfies
\([H_L,J_L]=0\).
To see this, expand the commutator as a sum of local terms.
Terms spanning four sites cancel by the Jacobi identity.
The remaining contribution is \(i\sum_xF(h_q)_x\), which vanishes
by the periodic sum of \eqref{eq:resh}.
Consequently, the equilibrium autocorrelation of \(J_L\) is
time independent. Its infinite-temperature value is determined
by the local traces below.

\subsection{Equilibrium current and energy fluctuations}
We evaluate the traces in the real representation
\eqref{eq:real}. Define
\begin{equation}
p(t)=5t^4-28t^3+51t^2-38t+64.
\end{equation}
The infinite-temperature susceptibilities are
\begin{equation}
\chi_J=\frac{\langle J_L^2\rangle_0}{L}=\frac49p(t),
\qquad
\chi_E=\frac{\langle H_L^2\rangle_0}{L}
=\frac43(t^2-2t+4).
\label{eq:chi}
\end{equation}
These expressions are independent of \(L\) for \(L\ge5\).
Their local contributions are
\begin{align}
\langle j_0^2\rangle_0
&=\frac49(3t^4-16t^3+33t^2-30t+64),\nonumber\\
\langle j_0j_1\rangle_0
&=\frac49t(t-4)(t-1)^2,\label{eq:contractions}\\
\langle h_0^2\rangle_0
&=\frac49(2t^2-4t+11),\qquad
\langle h_0h_1\rangle_0=\frac29(t-1)^2.\nonumber
\end{align}
All more distant contractions vanish. Thus each susceptibility
is the on-site contraction plus twice the adjacent one.
Appendix~\ref{app:transport} derives the traces by separating
the local current into terms with distinct spatial supports.

\begin{figure}[tb]
\centering
\begin{tikzpicture}
\begin{axis}[
width=.88\textwidth,height=5.8cm,
xlabel={$t=q^2$},ylabel={$\chi_J/(4\chi_E^2)$},
xmin=0,xmax=8,ymin=0,ymax=.4,
samples=241,domain=0:8,axis lines=left,
tick label style={font=\small},label style={font=\small},
grid=major,grid style={gray!15}]
\addplot[very thick,blue!65!black]
{(5*x^4-28*x^3+51*x^2-38*x+64)/(16*(x^2-2*x+4)^2)};
\addplot[only marks,mark=*,mark size=2pt,black]
coordinates {(0,.25) (1,.375) (4,.09375)};
\node[anchor=north,font=\small] at (axis cs:4,.07) {clock point};
\end{axis}
\end{tikzpicture}
\caption{Exact current variance with the energy scale removed:
\(\chi_J/(4\chi_E^2)=p(t)/[16(t^2-2t+4)^2]\).
The curve is evaluated directly from \eqref{eq:chi}.
Markers indicate \(t=0,1,4\), with \(t=4\) the clock point.
The rescaling separates variation of transport from the overall
growth of the local energy scale.}
\label{fig:transport}
\end{figure}
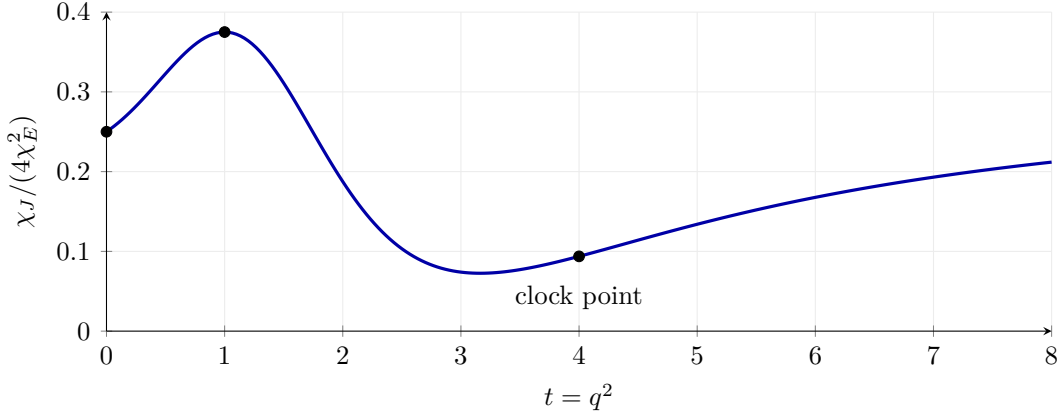

The adjacent-current contribution is essential to the coupling
dependence of \(\chi_J\). It survives at infinite temperature
because neighboring currents contain operators on a common bond.
It is negative for \(0<t<4\), except at its zero \(t=1\),
and positive for \(t>4\). A sum of individual local-current
norms would therefore overestimate the transport weight in
the first interval and underestimate it in the second.
Figure~\ref{fig:transport} displays the dimensionless ratio
\(\chi_J/(4\chi_E^2)\), which removes the overall local energy
scale and retains this dependence on the interaction.

\subsection{The high-temperature thermal Drude weight}
We use the conductivity convention
\(\operatorname{Re}\kappa(\nu)=
2\pi D_{\rm th}\delta(\nu)+\kappa_{\rm reg}(\nu)\).
For a conserved total energy current, the Kubo formula reduces to
\[
D_{\rm th}(\beta)=
\frac{\beta^2}{2}\lim_{L\to\infty}
\frac{\langle J_L^2\rangle_{\beta,c}}L,
\]
where the subscript \(c\) denotes the connected fluctuation.
Equation~\eqref{eq:chi} then gives
\begin{equation}
\lim_{\beta\downarrow0}\frac{D_{\rm th}(\beta)}{\beta^2}
=\frac29p(t)>0.
\label{eq:drude}
\end{equation}
The result fixes the leading high-temperature transport coefficient
throughout the real family.

At the clock point \(t=4\), \eqref{eq:drude} gives
\(\lim_{\beta\downarrow0}D_{\rm th}/\beta^2=48\).
This agrees with the nonchiral point of Ref.~\cite{mannadrude}.
For its convention \(J=f=1\), the high-temperature coefficient
is \(3\). After the on-site Fourier change of basis, the periodic
sum of \eqref{eq:clockpoint} equals \(-2H_{\rm clock}\).
Under \(H\mapsto aH\), the current scales as \(a^2\), so the
infinite-temperature Drude coefficient scales as \(a^4\):
\((-2)^4\times3=48\). The agreement fixes a common benchmark;
\eqref{eq:drude} gives the coupling dependence away from that point
in the family \eqref{eq:density}.

Current conservation expresses the connected fluctuation as
a pressure derivative with current source \(\lambda\):
\begin{equation}
\frac{\langle J_L^2\rangle_{\beta,c}}L
=\left.\partial_\lambda^2
\frac1L\log\Tr\exp(-\beta H_L+\lambda J_L)
\right|_{\lambda=0}.
\label{eq:sourcepressure}
\end{equation}
At fixed \(q\), both \(H_L\) and \(J_L\) are sums of bounded
finite-range operators. The high-temperature cluster expansion
gives a thermodynamic pressure analytic jointly near
\((\beta,\lambda)=(0,0)\), with locally uniform convergence of
its derivatives \cite[Corollary~33]{cluster}.
The limiting current fluctuation is therefore analytic near
\(\beta=0\), where its value is \eqref{eq:chi}.

Strict positivity follows from the decomposition
\begin{equation}
p(t)=(t-4)^2(t^2+2t+3)
+2(t-1)^2(2t^2-7t+8)>0,\qquad t\ge0.
\label{eq:positive}
\end{equation}
Both quadratic factors are positive, and the square factors
cannot vanish together.
The two terms arise from the orthogonal support decomposition
of the current given in Appendix~\ref{app:transport}.
They ensure that the total fluctuation stays nonzero even
when one contribution vanishes.
Together with analyticity, this proves a positive thermal
Drude weight throughout a sufficiently high-temperature
interval at each fixed \(q\).

\subsection{The energy-correlation second moment}
The same susceptibility also determines the spreading of
energy correlations. On the infinite chain, let
\[
C_E(x,s)=\langle h_x(s)h_0\rangle_0,\qquad
M_2(s)=\frac1{\chi_E}\sum_xx^2C_E(x,s).
\]
Stationarity and the continuity equation give
\[
\partial_s^2C_E(x,s)=C_J(x-1,s)-2C_J(x,s)+C_J(x+1,s),
\qquad C_J(x,s)=\langle j_x(s)j_0\rangle_0.
\]
Discrete summation by parts and conservation of the total
current imply \(M_2''(s)=2\chi_J/\chi_E\).
This is the density--current relation of Ref.~\cite{steinigeweg}
evaluated for the present chain.
Cyclicity of the trace gives \(C_E(x,s)=C_E(-x,-s)\), hence
\(M_2'(0)=0\), without a time-reversal assumption.
The adjacent energy-density contraction in
\eqref{eq:contractions} fixes \(M_2(0)\).
It follows that, at every finite time,
\begin{equation}
M_2(s)=\frac{(t-1)^2}{3(t^2-2t+4)}
+\frac{p(t)}{3(t^2-2t+4)}s^2.
\label{eq:moment}
\end{equation}
Locality controls the infinite-volume sums at fixed finite time.

The initial offset records the overlap of adjacent energy
densities, while the quadratic term gives their ballistic
spreading coefficient. Equation~\eqref{eq:moment} thus determines
the ballistic growth of the energy-correlation second moment.
The infinite-temperature calculation samples the full
three-state Hilbert space. To identify the degrees of freedom
governing the strong-coupling spectrum, we next examine
the ground space of the leading interaction.

\section{The strong-coupling limit of the open chain}\label{sec:open}
At large \(|q|\), the leading flag interaction leaves a
degenerate manifold on which the remaining terms act.
The effective spin chain is determined by resolving this manifold
before eliminating virtual excitations.
For an open chain in the real representation \eqref{eq:real},
write
\[
H_L(q)=q^2H_2+qH_1+H_0.
\]
Color \(0\) will be called the spectator and colors \(1,2\)
the active states. Spectator number labels the invariant sectors
of \(H_2\); the subleading terms couple these sectors.
The effective theories in this section and Sec.~\ref{sec:periodic}
are defined by \(|q|\to\infty\) at fixed \(L\).
Their thermodynamic quantities are obtained by subsequently
taking \(L\to\infty\).

\subsection{The full leading ground space}
For every \(L\ge2\), the ground energy of \(H_2\) is
\(E_2=-2(L-1)/3\), with degeneracy \(3\cdot2^{L-1}\).
The ground space contains a \(2^L\)-dimensional all-active
sector and a \(2^{L-1}\)-dimensional one-spectator sector.

The second sector arises from the spectator motion.
Let \(B\) be the signless incidence
matrix of a path, with columns \(b_j=e_j+e_{j+1}\).
For a fixed ordered word of active colors, the one-spectator
position Hamiltonian above \(E_2\) is \(BB^{\mathsf T}\).
Its unique zero mode is the alternating vector.
Consequently, each active word \(\alpha\) of length \(L-1\)
gives the ground state
\begin{equation}
\ket{g_\alpha}=\frac1{\sqrt L}\sum_{p=0}^{L-1}
(-1)^p\ket{p;\alpha}.
\label{eq:delocalized}
\end{equation}
Here \(\ket{p;\alpha}\) is obtained by inserting color \(0\)
after the first \(p\) letters of \(\alpha\).
The spectator is therefore delocalized over the whole chain.

For \(k\) spectators, the position Hamiltonian is the
\(k\)th exterior-power lift of \(BB^{\mathsf T}\).
Its eigenvalues are sums of \(k\) distinct one-particle
eigenvalues. There is only one zero eigenvalue, so two
or more spectators have strictly positive energy.
This establishes that the two sectors above exhaust the
leading ground space. The local action and the exterior-power
construction are given in Appendix~\ref{app:open}.

\subsection{Virtual transitions and the two XXZ Hamiltonians}
Let \(P_0\) project onto this full ground space.
The linear interaction replaces an equal active pair \(aa\)
by \(0\bar a\) or \(\bar a0\), where
\(\bar1=2\) and \(\bar2=1\).
The two possible spectator positions have opposite
amplitudes in \eqref{eq:delocalized}, giving
\(P_0H_1P_0=0\).
The first splitting is therefore of order \(q^0\):
the direct term \(H_0\) and the second-order virtual
transitions contribute at the same order.
The effective Hamiltonian is
\begin{equation}
H_{\rm eff}
=P_0H_0P_0-P_0H_1(H_2-E_2)^+H_1P_0,
\label{eq:heff}
\end{equation}
where the superscript \(+\) denotes the inverse on the
orthogonal complement of the ground space, extended by
zero on that space.

In the all-active sector the virtual process is governed by
\[
B^{\mathsf T}(BB^{\mathsf T})^+B=\I.
\]
It subtracts
\(\Pi^{\rm equal}=(\I+ZZ)/2\) on each active bond.
Combining this term with the direct compression
\(2\I/3+XX+YY\) gives the spin-\(1/2\) XXZ Hamiltonian
\begin{equation}
K_m=\sum_{j=1}^{m-1}
\left(X_jX_{j+1}+Y_jY_{j+1}-\frac12Z_jZ_{j+1}
+\frac16\I\right),
\label{eq:K}
\end{equation}
where \(X,Y,Z\) are Pauli matrices on the active colors.
Thus the anisotropy \(\Delta=-1/2\) includes a contribution
from virtual spectator states.

In the one-spectator sector, the intermediate states
contain two spectators. Evaluating their resolvent gives
a factor \(n/(n+1)\), where \(n=L-1\).
The direct term has the same factor multiplying the
active bonds, together with a scalar contribution from
the bonds touching the spectator.
The two blocks of \eqref{eq:heff}, in the zero-spectator
and one-spectator sectors respectively, are
\begin{equation}
K_L,\qquad
\frac{10n}{3(n+1)}\I+\frac n{n+1}K_n,\qquad n=L-1.
\label{eq:blocks}
\end{equation}
Appendix~\ref{app:open} derives the two-spectator inverse
and both normalizations.
The sectors do not mix at this order.
The corresponding low-energy eigenvalues
of the original Hamiltonian are
\(q^2E_2+\spec H_{\rm eff}+O(|q|^{-1})\).

\subsection{Spectator energy cost and XXZ thermodynamics}
The relative scale and scalar offset in \eqref{eq:blocks}
allow a comparison of the two effective ground energies.
Let \(e_m=\min\spec K_m\).
A product-state bound gives \(e_m\le0\), and adjoining a
maximally mixed site gives \(e_{m+1}\le e_m+1/6\).
It follows that
\begin{equation}
\frac{10n}{3(n+1)}+\frac n{n+1}e_n-e_{n+1}
\ge\frac{19n-1}{6(n+1)}>0.
\label{eq:gap}
\end{equation}
The effective ground state therefore lies in the all-active
block for every length. The spectator sector, despite
belonging to the leading ground space, has a finite positive
energy cost after the degeneracy is lifted.

In the normalization of \eqref{eq:K}, the XXZ solution
\cite{yangyang,xxzboundary,xxzcft} gives
\begin{equation}
\varepsilon_K=\lim_{m\to\infty}\frac{e_m}{m}
=\frac53-\frac{3\sqrt3}{2},\qquad
c=1,\quad K_{\rm LL}=\frac32,\quad v_{\rm XXZ}=\frac{3\sqrt3}{2}.
\label{eq:xxzdata}
\end{equation}
Here \(K_{\rm LL}\) is the Luttinger parameter and
\(v_{\rm XXZ}\) is the velocity in the effective spin chain.
The ground-energy integral in this normalization is
evaluated in Appendix~\ref{app:open}.
Using the open-chain expansion
\(e_m=(m-1)\varepsilon_K+e_s+o(1)\), the boundary constant
\(e_s\) cancels from the difference between the two block
minima, yielding
\begin{equation}
\lim_{n\to\infty}
\left(\frac{10n}{3(n+1)}+\frac n{n+1}e_n-e_{n+1}\right)
=\frac{10}{3}-2\varepsilon_K=3\sqrt3.
\label{eq:defect}
\end{equation}

The uniform positive bound \eqref{eq:gap} selects the critical
XXZ sector, while \eqref{eq:defect} fixes the spectator excitation
energy in the thermodynamic limit.
We next examine how closing the chain changes the virtual
process that produced \eqref{eq:K}.

\section{Periodic chains and the XXZ supercharge}\label{sec:periodic}
The open-chain XXZ interaction followed from the virtual
propagation of a spectator on a configuration path.
Closing the chain changes this propagation and leaves an
additional term in the effective Hamiltonian.
We consider a ring of length \(L\ge3\) in the uniform real
gauge. As explained in Sec.~\ref{sec:model}, this is the
original untwisted three-state chain when \(3\mid L\).

\subsection{Configuration cycles and the virtual correction}
With one spectator, the ordered active colors form a cyclic
binary word, or necklace. If its minimal rotational period
is \(p\mid L-1\), the spectator must complete \(p\) circuits
of the physical ring to recover the original configuration.
The position graph is therefore a cycle with \(Lp\) vertices.
An alternating one-spectator zero mode exists exactly when
\(Lp\) is even. The full leading ground space consists of
these modes and the all-active states; its dimension is
given in \eqref{eq:necklaces} in Appendix~\ref{app:periodic}.

\begin{figure}[tb]
\centering
\begin{minipage}[t]{.48\textwidth}
\centering
\begin{tikzpicture}[font=\small]
\path[use as bounding box] (-.6,-1.95) rectangle (5.4,1.9);
\node at (2.4,1.55) {(a) Open configuration path};
\foreach \i/\s in {0/+,1/-,2/+,3/-,4/+,5/-}{
  \node[circle,draw=blue!65!black,fill=blue!12,
    minimum size=5pt,inner sep=0pt,label=above:{$\s$}]
    (p\i) at (.96*\i,0) {};
}
\foreach \i/\j in {0/1,1/2,2/3,3/4,4/5}{
  \draw[thick,blue!65!black] (p\i)--(p\j);
}
\node at (2.4,-.75) {$\ker B=\{0\}$};
\node at (2.4,-1.55) {$B^\dagger(BB^\dagger)^+B=\I$};
\end{tikzpicture}
\end{minipage}\hfill
\begin{minipage}[t]{.48\textwidth}
\centering
\begin{tikzpicture}[font=\small]
\path[use as bounding box] (-3,-1.95) rectangle (3,1.9);
\node at (0,1.55) {(b) Even configuration cycle};
\foreach \i/\angle in {0/0,1/60,2/120,3/180,4/240,5/300}{
  \node[circle,draw=blue!65!black,fill=blue!12,
    minimum size=5pt,inner sep=0pt] (c\i) at (\angle:1cm) {};
}
\foreach \i/\j/\s in {0/1/+,1/2/-,2/3/+,3/4/-,4/5/+,5/0/-}{
  \draw[thick,blue!65!black] (c\i)--node[midway,fill=white,
    inner sep=1.2pt,text=red!65!black] {$\s$} (c\j);
}
\node at (0,-1.55) {$B^\dagger(BB^\dagger)^+B=\I-\Pi_{\rm alt}$};
\end{tikzpicture}
\end{minipage}
\caption{Configuration graphs governing virtual spectator motion.
Vertices are one-spectator configurations, and edges label pair-contraction
events; \(B\) maps edge amplitudes to vertex amplitudes.
The path has an alternating vertex zero mode (signs in panel a) but no
edge kernel. An even configuration cycle also has an alternating edge
kernel (signs in panel b). The missing virtual subtraction is consequently
\(C_L=F_L^\dagger\Pi_{\rm alt}F_L\), identified in
\eqref{eq:supercharge} with a supercharge square.
The displayed graphs are schematic: a periodic configuration cycle
has length \(Lp\), where \(p\) is the active necklace period.
OpenAI Codex (OpenAI, model \texttt{gpt-6-astra}) assisted with
writing the TikZ code for this schematic.}
\label{fig:virtual}
\end{figure}
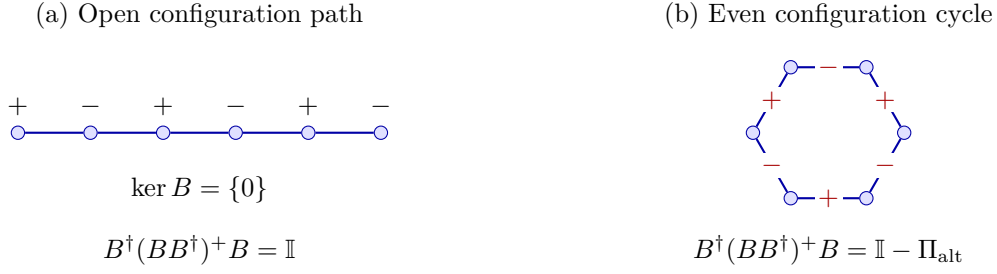

The distinction from an open configuration path also appears
in the space of contraction events.
On an even cycle, the signless incidence matrix \(B\) has
an alternating edge kernel, so the identity used in the
open-chain calculation becomes
\[
B^\dagger(BB^\dagger)^+B=\I-\Pi_{\rm alt}.
\]
Here \(\Pi_{\rm alt}\) projects onto the alternating edge
vectors of the even configuration cycles; odd cycles have
no such kernel.
The virtual subtraction is reduced by a positive operator
\(C_L=F_L^\dagger\Pi_{\rm alt}F_L\), where \(F_L\) records
the labeled contractions of equal active pairs.
Thus the extra effective interaction is fixed by the
change in the spectator configuration graph.
Figure~\ref{fig:virtual} illustrates this mechanism.

\subsection{Identification with the XXZ supercharge}
The dynamical supersymmetry of the \(\Delta=-1/2\) XXZ chain
\cite{hagendorf} identifies \(C_L\) as a supercharge square.
Let \(T_N\) be right translation on an \(N\)-site active
chain. The length-changing supercharge \(Q_N\) acts on the
sector \(T_N=(-1)^{N+1}\) and maps it to the corresponding
sector at length \(N+1\).
It is generated by the local map
\(\mathsf s\ket b=\ket{\bar b\,\bar b}/\sqrt2\), with
the active colors relabeled \(b=0,1\) and \(\bar b=1-b\).
The normalized signed insertion sum defining \(Q_N\)
is given in Appendix~\ref{app:periodic}.

With these conventions, the all-active second-order
effective Hamiltonian is
\begin{equation}
H_{\rm eff,per}^{(0)}
=K_L^{\rm per}+C_L,\qquad
C_L=\frac2{L-1}Q_{L-1}Q_{L-1}^\dagger.
\label{eq:supercharge}
\end{equation}
Here \(K_L^{\rm per}\) is the periodic sum of the XXZ
density in \eqref{eq:K}.
The coefficient \(2/(L-1)\) follows by matching the
normalized necklace states to the signed insertion events.
This matching, including necklaces with shorter rotational
periods, is established in \eqref{eq:necklaceimage}.

The correction is supported in \(T_L=(-1)^{L+1}\) and obeys
\begin{equation}
[K_L^{\rm per},C_L]=0,\qquad
0\le C_L,\qquad \|C_L\|\le\frac{3L}{L-1}.
\label{eq:Cbound}
\end{equation}
The commutation relation follows from the nilpotence
\(Q_{N+1}Q_N=0\) and the supersymmetric representation
of the XXZ Hamiltonian.
Appendix~\ref{app:periodic} gives the cancellation of
insertion terms and the local estimate yielding the norm
bound. Nilpotence also determines the spectral support
of the three-state virtual correction.

\subsection{Spectral support of the correction}
The supercharge relation determines which XXZ levels can shift.
Write \(e_m^{\rm per}=\min\spec K_m^{\rm per}\).
Nilpotence gives the adjacent-length intertwining relation
\eqref{eq:adjacentlength}: the energies carried by
\(\operatorname{im}Q_{L-1}\) are XXZ energies at length \(L-1\)
shifted by \(5/3\). On this image, a level of energy \(E\)
acquires the correction \((5L/3-E)/(L-1)\).
Every affected level therefore satisfies
\(E\ge e_{L-1}^{\rm per}+5/3\).

Inserting a maximally mixed spin into a translation-averaged
XXZ ground state gives
\(e_L^{\rm per}\le (L-2)e_{L-1}^{\rm per}/(L-1)+1/3\).
Combining this variational estimate with a three-site lower
bound separates the affected levels from the ground energy:
\[
e_{L-1}^{\rm per}+\frac53-e_L^{\rm per}
\ge\frac{7-\sqrt{33}}4>0,\qquad L\ge4.
\]
Appendix~\ref{app:periodic} gives both estimates explicitly.
Thus \(C_L\) vanishes on a length-independent interval above
the XXZ ground state. The all-active block has ground energy
\(e_L^{\rm per}\). At \(L=3\), the common ground energy is
\(-1\), with degeneracy six for XXZ and five for the effective
block; Appendix~\ref{app:periodic} gives the rank-one splitting.

\subsection{Periodic ground-state selection}
The remaining leading ground states contain one spectator.
With \(n=L-1\), their alternating insertion map
\eqref{eq:delocalized} is an isometry from the active momentum
sector \(\Pi_n\), where \(T_n=(-1)^{n+1}\).
Deleting any one bond of the ring gives an open chain whose
ground space contains these states.
A matrix Cauchy--Schwarz inequality compares the periodic
two-spectator resolvent with the sum of the open-chain
resolvents. Using the open block \eqref{eq:blocks} yields
the operator bound
\[
H_{\rm eff,per}^{(1)}\ge\frac{10}{3}\I+
\frac{n-1}{n}\left.K_n^{\rm per}\right|_{\Pi_n}.
\]
The construction and the factors from the bond cuts are
derived in \eqref{eq:cutresolvent}--\eqref{eq:periodicspectatorbound}.
Together with the XXZ variational estimate, this implies
\begin{equation}
\min\spec H_{\rm eff,per}^{(1)}
-\min\spec H_{\rm eff,per}^{(0)}\ge3,\qquad L\ge3.
\label{eq:periodicselection}
\end{equation}
The all-active sector is therefore the ground block of the
complete periodic effective Hamiltonian.

Combining the spectator bound with the support of \(C_L\)
gives, with multiplicities,
\begin{equation}
\spec_{<e_L^{\rm per}+\delta_*}H_{\rm eff,per}
=\spec_{<e_L^{\rm per}+\delta_*}K_L^{\rm per},\qquad
\delta_*=\frac{7-\sqrt{33}}4,\quad L\ge4.
\label{eq:fullperiodiclowenergy}
\end{equation}
Here the subscript restricts the spectrum to energies below
the indicated threshold. In particular, the ground-state
multiplicity and all excitations with energy less than
\(\delta_*\) above the ground state agree exactly.
The conformal \(1/L\) spectrum of the periodic effective
theory is consequently the XXZ spectrum with the parameters
in \eqref{eq:xxzdata}.

At any fixed positive temperature, the operator bound also
controls the spectator partition function by that of an
\(n\)-site XXZ chain with inverse temperature
\(\beta(n-1)/n\). Together with \eqref{eq:Cbound}, this
proves that the full periodic effective Hamiltonian has
the XXZ bulk free-energy density.

\section{Discussion}
The three-state chain has a conserved total energy current
at every real coupling,with the exact leading high-temperature coefficient of the thermal Drude weight \eqref{eq:drude}.
Adjacent-current correlations control its coupling dependence:
they suppress the current variance for \(0<t<4\), except at
\(t=1\), and enhance it for \(t>4\).
The positive decomposition \eqref{eq:positive} proves a
nonzero ballistic contribution throughout the real family
at sufficiently high temperature.
The same current variance fixes the quadratic growth of the
energy-correlation second moment \eqref{eq:moment}.

In the strong-coupling effective theory, virtual spectator
transitions generate the XXZ anisotropy \(\Delta=-1/2\)
and select the all-active ground sector.
The open-chain calculation resolves the entire flag ground
space into two explicitly normalized XXZ blocks.
Their ground-energy difference is positive at every length
and tends to \(3\sqrt3\).
The third local state therefore participates in the virtual
process that produces the exchange interaction and acquires
a finite excitation energy when the degeneracy is lifted.

Closing the chain converts the configuration path into a
cycle. Its alternating edge kernel generates the positive
supercharge square \(C_L\), with coefficient and momentum
support determined by the three-state embedding.
The adjacent-length supercharge relation places the entire
support of \(C_L\) above the protected XXZ energy interval.
The resolvent bound independently separates the spectator
block from the ground state by at least three energy units.
Together they establish \eqref{eq:fullperiodiclowenergy}:
the full periodic effective Hamiltonian has the exact XXZ
low-energy spectrum, including ground-state degeneracy and
the conformal \(1/L\) levels.
The partition-function bound extends the XXZ correspondence
to the bulk free-energy density at every positive temperature.

Two further problems follow from this structure.
Current conservation reduces the finite-temperature Drude
weight to the equilibrium fluctuation of \(J_L\), with
\eqref{eq:drude} supplying its exact high-temperature limit.
The finite-coupling spectral problem centers on the joint
\(q,L\) dependence of the corrections generated by the
spectator resolvent as its leading position gap closes.
\appendix
\section*{Appendix}
\section{The regular R-matrix}\label{app:spectral}
This appendix specifies the \(R\)-matrix underlying the local
conservation identity \eqref{eq:resh}.
We give its entries in the charge-sector ordering of
\eqref{eq:density}, establish the Hamiltonian normalization and
describe the polynomial verification of the Yang--Baxter equation.
The construction starts from the regular expansion
\(\check R(u)=\I+u h_q+O(u^2)\).
Differentiating \eqref{eq:yb} in one additive parameter at zero gives
the Sutherland relation
\[
\check R_{12}\check R'_{23}-\check R'_{12}\check R_{23}
=h_{23}\check R_{12}\check R_{23}
-\check R_{12}\check R_{23}h_{12}.
\]
All spectral matrices in this relation are evaluated at \(u\).
Solving its coefficient equations with \(\tr\check R=9\)
generates the Taylor expansion. Resumming it with
\(x=e^{i\sqrt3u}\) and removing a scalar factor gives the
degree-four polynomial solution displayed below.
The final verification uses the full Yang--Baxter equation.
The changes of basis are constant in both \(q\) and \(x\):
\[
U_0=\begin{pmatrix}1&0&0\\0&1&-\omega^2\\0&\omega&1\end{pmatrix},
\quad U_1=\begin{pmatrix}1&0&-\omega^2\\\omega&0&1\\0&1&0\end{pmatrix},
\quad U_2=\begin{pmatrix}1&0&-\omega\\0&1&0\\\omega^2&0&1\end{pmatrix}.
\]
Write \(\kappa=1+2\omega=i\sqrt3\) and
\begin{align*}
a(x)&=(x-1)^2(x^2-x+1),&
b(x)&=x^4-8x^2+1,\\
N_q(x)&=b(x)-q^2a(x),&
F(x)&=x(x^2-1),\\
K_q(x)&=(x^2-1)\{q^2(x^2-x+1)-(x^2+1)\}.&&
\end{align*}
Define the two-dimensional blocks
\[
A_0(x)=
\begin{pmatrix}
-(\kappa K_q+N_q)/6&-2\kappa F/3\\
-\kappa F/3&(\kappa K_q-N_q)/6
\end{pmatrix},\qquad
A_1(x)=\begin{pmatrix}\alpha_q(x)&\beta_q(x)\\
2\beta_q(x)&\delta_q(x)\end{pmatrix},
\]
where
\begin{align*}
\alpha_q(x)&=-\frac{\omega}{3}(x+\omega^2)(x-\omega)
\bigl[q^2\{x^2+(\omega-1)x-\omega\}-x^2+\omega\bigr],\\
\beta_q(x)&=\frac{1-\omega}{3}qx(x-1)(x+\omega^2),\\
\delta_q(x)&=-\frac13(x+\omega^2)(x-\omega^2)
\bigl[q^2\{x^2+(\omega^2-1)x-\omega^2\}-x^2+\omega^2\bigr].
\end{align*}
The scalar channels are
\begin{align*}
\Phi_q(x)&=q^2(x^2+\omega^2x+\omega)-x^2+2\omega^2x-\omega,\\
c_0(x)&=\frac{\omega^2}{3}(x+\omega^2)^2\Phi_q(x),&
c_1(x)&=\frac{\omega}{3}(x+\omega^2)(x+\omega)\Phi_q(x).
\end{align*}
The required matrix is
\begin{equation}
\begin{aligned}
U_0^{-1}\mathcal R_q(x)|_{\mathcal H_0}U_0&=A_0(x)\oplus c_0(x),\\
U_1^{-1}\mathcal R_q(x)|_{\mathcal H_1}U_1&=A_1(x)\oplus c_1(x),\\
U_2^{-1}\mathcal R_q(x)|_{\mathcal H_2}U_2&=
D A_1(x)D^{-1}\oplus c_1(x),\qquad
D=\operatorname{diag}(\omega,1).
\end{aligned}
\label{eq:closedR}
\end{equation}
In particular, all entries are polynomials in \(q,x\) with coefficients
in \(\C\). Substitution of \(x=1\), using
\(\omega^2+\omega+1=0\), gives
\[
A_0(1)=A_1(1)=\I_2,\qquad c_0(1)=c_1(1)=1.
\]
Differentiating \eqref{eq:closedR} gives
\[
i\sqrt3\left(\mathcal R_q'(1)
-\frac{\tr\mathcal R_q'(1)}9\I\right)=h_q,
\]
which proves the normalization in \eqref{eq:regular}.

Direct substitution of these polynomials proves the
Yang--Baxter equation. If
\(w_{ab}^{cd}(x)=\langle cd|\mathcal R_q(x)|ab\rangle\),
then \(w_{ab}^{cd}=0\) unless \(a+b=c+d\pmod3\).
For fixed input \(a,b,c\) and output \(d,e,f\), the identity is
\begin{equation}
\sum_{r,s,k}
w_{ab}^{rs}(y)w_{sc}^{kf}(xy)w_{rk}^{de}(x)
=
\sum_{r,s,k}
w_{bc}^{rs}(x)w_{ar}^{dk}(xy)w_{ks}^{ef}(y).
\label{eq:weightidentity}
\end{equation}
When the total charges disagree both sides vanish. Otherwise, on the
left set \(s=a+b-r\), \(k=a+b+c-f-r\), and on the right set
\(s=b+c-r\), \(k=a+r-d\), all modulo three. Each side then has only
three terms. Inserting \eqref{eq:closedR} and collecting powers of
\(q,x,y\) gives equal coefficients after reduction by
\(\omega^2+\omega+1=0\). These coefficient identities establish
\eqref{eq:yb} for arbitrary spectral parameters.
The accompanying script \texttt{verify\_symbolic.py} checks all 243
charge-compatible components over
\(\mathbb Q[q,x,y,\omega]/(\omega^2+\omega+1)\), as well as
regularity and the Hamiltonian normalization. The other 486
components vanish by charge conservation.

For a compact certificate of the energy-current identity, write
\(R_k=\partial_u^k\check R_q(0)\), using the trace normalization
in \eqref{eq:regular}. The quadratic coefficient equations give
\(R_2=h_q^2-\tr(h_q^2)\I/9\). The coefficient of \(u^2v\)
then gives \eqref{eq:resh} with
\begin{equation}
Y=R_3-h_q^3+\frac{\tr(h_q^2)}3h_q.
\label{eq:Ycertificate}
\end{equation}
This specifies \(Y\) directly in terms of the displayed polynomial
matrix. The arXiv ancillary verification files also evaluate \eqref{eq:Ycertificate}
and checks \(F(h_q)=Y_{23}-Y_{12}\) by exact matrix multiplication.

At real \(q,x\), the matrix satisfies the scalar unitarity identity
\begin{equation}
\mathcal R_q(x)^\dagger\mathcal R_q(x)=g_t(x)\I,\qquad
g_t(x)=\frac{(x^2-x+1)^2}{9}p_t(x),
\end{equation}
where
\begin{align*}
p_t(x)&=(t-1)^2(x^4+1)-(t-1)(t+2)(x^3+x)+3(2t+1)x^2\\
&=\left((t-1)x^2-\frac{t+2}{2}x+\frac{1-t}{2}\right)^2
+\frac34\left((t-1)-(t+2)x\right)^2 .
\end{align*}
For \(t\ge0\), the two squares cannot vanish simultaneously at \(x>0\).
For \(t\le1\), the root of the second factor is not positive.
For \(t>1\), substituting \(x=(t-1)/(t+2)\) into the first factor
gives \(-3(t-1)(2t+1)/(t+2)^2\ne0\).
The positive real spectral axis therefore has a nonsingular unitary
normalization.

\subsection{Computational methods}\label{app:methods}
The arXiv ancillary checks use Python and SymPy with exact polynomial
and rational arithmetic. The spectral and transport scripts
independently transcribe the displayed matrices; the latter
evaluates local traces and the derived transport coefficients.
OpenAI Codex (OpenAI), using models \texttt{gpt-6-astra} and
\texttt{gpt-5.6-sol}, assisted with literature searches, algebraic
manipulations and proof development. Model \texttt{gpt-6-astra} also assisted with
developing the verification scripts and figure code.
Figure~\ref{fig:transport} is rendered by PGFPlots from
\eqref{eq:chi}, using the expression and plotting parameters
included in the LaTeX source. Figure~\ref{fig:virtual} is a TikZ
schematic of the incidence relations derived in
Appendices~\ref{app:open} and \ref{app:periodic}.
\section{Local trace calculation}\label{app:transport}
We evaluate the contractions entering \eqref{eq:chi} and
identify the two nonnegative contributions to the current
variance. The useful step is to separate the three-site current
according to the support of its operator components.
All traces in this appendix are normalized by the local Hilbert-space
dimension, unless \(\tr\) is written explicitly for a \(3\times3\)
matrix. The real form \eqref{eq:real} is valid on each local interval.
For example, the coefficients of \(e_i\otimes e_j^\dagger\) in the
original density form the matrix \(\omega DND^\dagger\), with
\(D=\operatorname{diag}(1,\omega^2,\omega)\).
Conjugation by \(S\) removes \(D\), and the relative power of \(Z\) on
adjacent sites removes \(\omega\).

Put \(d_i=v_{i+1,i+1}-v_{ii}\), so
\((d_0,d_1,d_2)=(1-t,0,t-1)\).
The three-site current decomposes as
\begin{equation}
j=J^{(3)}+E_{12}+G_{23},\qquad
J^{(3)}=i[\mathcal T_{12},\mathcal T_{23}],\quad
E=i[\mathcal T,\I\otimes v],\quad
G=i[v\otimes\I,\mathcal T].
\label{eq:orthogonal}
\end{equation}
Every one-site partial trace of \(J^{(3)}\) is zero. Both partial
traces of \(E\) and \(G\) vanish. These three terms are therefore
orthogonal in Hilbert--Schmidt inner product. Two adjacent currents
have only one surviving contraction, between \(G\) and \(E\) on
their common bond:
\begin{equation}
\langle j_0^2\rangle_0=
\langle(J^{(3)})^2\rangle_0+\langle E^2\rangle_0+\langle G^2\rangle_0,
\qquad
\langle j_0j_1\rangle_0=\langle GE\rangle_0.
\end{equation}
At greater separation a nonoverlapping end site has zero partial
trace, so the contraction vanishes.

Orthogonality of matrix units gives
\begin{align}
\langle E^2\rangle_0=\langle G^2\rangle_0
&=\frac29\sum_{i,j}N_{ij}^2d_j^2
=\frac49(t-1)^2(t^2-3t+8),\\
\langle GE\rangle_0
&=\frac29\sum_{i,j}N_{ij}^2d_id_j
=\frac49t(t-4)(t-1)^2.
\end{align}
For the three-site term, the only commutators needed are
\[
[e_j^\dagger,e_k]=\delta_{jk}(E_{jj}-E_{j+1,j+1}),
\qquad
[e_j,e_k]=-\sum_m\varepsilon_{jkm}e_m^\dagger.
\]
Split \([\mathcal T_{12},\mathcal T_{23}]\) into four orthogonal parts
according to the raising or lowering operator at each end.
For end indices \(i,\ell\), the mixed parts have middle factor
\[
D_{i\ell}=\sum_jN_{ij}N_{j\ell}(E_{jj}-E_{j+1,j+1}),
\]
while the equal-direction parts are the cross products of rows
\(i,\ell\) of \(N\). If \(n_j=(N^2)_{jj}\), the cyclic-difference
and cross-product norm identities give
\begin{align*}
\sum_{i,\ell}\|D_{i\ell}\|_{\rm HS}^2
&=3\sum_jn_j^2-\tr N^4,\\
\sum_{i,\ell}\|N_{i\bullet}\times N_{\ell\bullet}\|^2
&=(\tr N^2)^2-\tr N^4.
\end{align*}
Consequently,
\begin{equation}
\langle(J^{(3)})^2\rangle_0
=\frac2{27}\left(3\sum_jn_j^2+(\tr N^2)^2-2\tr N^4\right)
=\frac49(t-4)^2(t^2+2t+3).
\label{eq:threepoint}
\end{equation}
The last equality uses only
\[
n_0=n_2=t^2-3t+8,\qquad n_1=2t+4,\qquad
\spec N=\{0,t+2,t-4\}.
\]
The antisymmetric vector \((1,0,-1)\) has eigenvalue \(t-4\);
the complementary \(2\times2\) block has determinant
\(2t-2q^2=0\) and trace \(t+2\).

The energy terms similarly satisfy
\[
\langle h_0^2\rangle_0
=\frac23\tr v^2+\frac29\tr N^2,\qquad
\langle h_0h_1\rangle_0=\frac13\tr v^2.
\]
These formulas produce all four contractions in
\eqref{eq:contractions}. Taking the on-site value plus twice the
neighboring value yields \eqref{eq:chi}. Equivalently,
\[
\chi_J=\langle(J^{(3)})^2\rangle_0+\langle(E+G)^2\rangle_0,
\]
which directly explains the two nonnegative terms in
\eqref{eq:positive}.

The arXiv ancillary script \texttt{verify\_transport.py} reconstructs
\eqref{eq:real} from its matrix elements independently of the
spectral verification script. It checks the local contractions,
\eqref{eq:chi}, the positivity decomposition and both coefficients
in \eqref{eq:moment} by exact symbolic arithmetic, including the
clock-point normalization comparison.
\section{Derivation of the open-chain effective Hamiltonian}\label{app:open}
We first identify the complete ground space of the leading
interaction. The one-spectator and two-spectator resolvents
then determine the two effective blocks in \eqref{eq:blocks}.

\subsection{Local action and leading ground space}
Expand \eqref{eq:real} as
\(\widetilde h_q=q^2h^{(2)}+qh^{(1)}+h^{(0)}\).
The leading local term is positive after a scalar shift:
\begin{equation}
h^{(2)}+\frac23\I
=2\ket{00}\bra{00}
+\sum_{a=1,2}(\ket{0a}+\ket{a0})(\bra{0a}+\bra{a0}).
\label{eq:flag}
\end{equation}
The nonzero actions of the linear term are
\begin{equation}
h^{(1)}\ket{aa}=\ket{0\bar a}+\ket{\bar a0},\qquad
h^{(1)}\ket{0b}=h^{(1)}\ket{b0}=\ket{\bar b\,\bar b}.
\label{eq:collapse}
\end{equation}
The active compression of \(h^{(0)}\) and its mixed block are
\begin{equation}
h^{(0)}_{\rm act}=\frac23\I+XX+YY,\qquad
h^{(0)}_{\{\ket{0a},\ket{a0}\}}
=\begin{pmatrix}-1/3&-2\\-2&-1/3\end{pmatrix}.
\label{eq:constant}
\end{equation}
The remaining spectator-changing entries connect \(00\) with \(12,21\).
Thus \(H_0\) preserves spectator-number parity.

Let \(B\) be the \(L\times(L-1)\) signless incidence matrix of a path,
with columns \(b_j=e_j+e_{j+1}\), \(0\le j\le L-2\), and put
\(Q=BB^{\mathsf T}\). Its kernel is the alternating vector
\(a=\sum_{p=0}^{L-1}(-1)^pe_p\), and all other eigenvalues are positive.
Fix the ordered active word left after deleting \(k\) spectators.
Their positions \(p_1<\cdots<p_k\) identify the position space with
\(e_{p_1}\wedge\cdots\wedge e_{p_k}\). Equation~\eqref{eq:flag}
then gives
\begin{equation}
A_L:=H_2+\frac23(L-1)\I=d\Gamma_k(Q).
\label{eq:exterior}
\end{equation}
Here \(d\Gamma_k(Q)\) is the sum of \(Q\) acting on the \(k\) exterior
factors. Each adjacent move \(0a\leftrightarrow a0\) has coefficient
one and does not cross another spectator, so no exterior sign is
introduced. A \(00\) bond contributes two diagonal units, one for
each occupied endpoint, as does the sum of position degrees.

The spectrum in \eqref{eq:exterior} consists of sums of \(k\)
distinct eigenvalues of \(Q\). There is only one zero eigenvalue.
Thus \(k\ge2\) has strictly positive energy, \(k=0\) has \(2^L\)
zero modes, and \(k=1\) has one alternating mode
\eqref{eq:delocalized} for each length-\(L-1\) active word.
This proves the leading energy and the full ground-space
decomposition used in Sec.~\ref{sec:open}.

\subsection{Elimination of virtual states in the all-active block}
Each equal pair \(aa\) contracts to \(\bar a\), with a spectator
inserted on either side. For a fixed contracted active word the
two positions are exactly the column \(b_j\).
Let \(F\) record these contractions with their edge labels.
The transition map is \(BF\). Since \(B^\dagger a=0\), it is
orthogonal to every one-spectator zero mode and
\(P_0H_1P_0=0\).
The path incidence matrix has full column rank, so
\[
B^{\mathsf T}(BB^{\mathsf T})^+B=\I .
\]
Expanding a labeled contracted color recovers the initial active
word uniquely. Therefore
\begin{equation}
P^{(0)}H_1A_L^+H_1P^{(0)}
=F^{\mathsf T}F
=\sum_{j=1}^{L-1}\Pi^{\rm equal}_{j,j+1},\qquad
\Pi^{\rm equal}=\frac12(\I+ZZ).
\end{equation}
Subtracting this from the first formula in \eqref{eq:constant}
gives \(K_L\).

\subsection{Two-spectator virtual states}
In a one-spectator ground state, the two positions that can annihilate
the existing spectator have opposite amplitudes. Their images under
\(H_1\) cancel. The remaining transitions contract an equal active
pair away from the existing spectator, and hence contain two
spectators.

Fix the active word after that contraction and the contraction
position \(0\le j\le L-3\). Define position vectors
\[
u_j=\sum_{p=0}^{j}(-1)^pe_p,\qquad
v_j=\sum_{r=j+2}^{L-1}(-1)^{r-j-2}e_r,\qquad
z_j=\frac1{\sqrt L}u_j\wedge v_j.
\]
Adjacent terms cancel in the path matrix:
\[
Qu_j=(-1)^jb_j,\qquad Qv_j=b_{j+1}.
\]
The exact inverse image of the virtual transition is therefore
\begin{equation}
y_j=d\Gamma_2(Q)z_j
=\frac1{\sqrt L}
\bigl((-1)^jb_j\wedge v_j+u_j\wedge b_{j+1}\bigr).
\label{eq:inverse}
\end{equation}
The second term lists the original spectator positions to the left
of the contracted pair, and the first lists those to its right.
If the spectator is between the pair, that pair is not adjacent
and cannot contract.

Exterior inner products are \(2\times2\) Gram determinants.
For \(i<j\), \(u_i\) is disjoint from \(b_j,v_j,b_{j+1}\),
and the pairing of \(v_i\) with \(b_{j+1}\) cancels between its
two alternating components. Thus
\[
\langle z_i,y_j\rangle=0\quad(i\ne j),\qquad
\langle z_j,y_j\rangle
=\frac{\|u_j\|^2+\|v_j\|^2}{L}=\frac{L-1}{L}.
\]
Self-adjointness handles \(i>j\).
Positive definiteness of \(d\Gamma_2(Q)\) makes
\(z_j=d\Gamma_2(Q)^{-1}y_j\) the unique solution of
\eqref{eq:inverse}.
Different contraction positions have zero cross terms, while a
fixed position and contracted word determine the original word.
With \(n=L-1\), the full virtual term is
\begin{equation}
P^{(1)}H_1A_L^+H_1P^{(1)}
=\frac n{n+1}\sum_{j=1}^{n-1}\Pi^{\rm equal}_{j,j+1}.
\label{eq:virtualone}
\end{equation}

\subsection{Direct term and perturbation estimate}
An adjacent pair in an active word of length \(n\) remains physically
adjacent at \(n\) of the \(n+1\) possible spectator insertion positions.
The active direct term is therefore
\[
\frac n{n+1}\sum_{j=1}^{n-1}
\left(\frac23\I+X_jX_{j+1}+Y_jY_{j+1}\right).
\]
The mixed bonds give the position operator
\(-D_{\rm path}/3-2A_{\rm path}\), where the two matrices are the
degree and adjacency matrices of the path. Its expectation in the
alternating vector uses
\[
a^{\mathsf T}D_{\rm path}a=2n,\qquad
a^{\mathsf T}A_{\rm path}a=-2n,\qquad a^{\mathsf T}a=n+1.
\]
This supplies \(10n/[3(n+1)]\).
Subtracting \eqref{eq:virtualone} gives the second block of
\eqref{eq:blocks}. The two ground sectors do not mix:
\(H_0\) preserves spectator parity, and the \(H_1\) images of the
two ground sectors have respectively one and two spectators.

For fixed \(L\), \(A_L\) is gapped on the orthogonal complement of
the ground space. Apply the block eigenvalue equation to
\(H_L/q^2=H_2+q^{-1}H_1+q^{-2}H_0\).
The excited component is, to first order,
\(-q^{-1}A_L^+H_1\) times the ground component.
Substitution gives \eqref{eq:heff}, with remainder
\(O(|q|^{-1})\) in the original energy units.
The variational argument \eqref{eq:gap} then proves selection of
the zero-spectator effective ground block.

\subsection{Ground-energy integral}\label{app:xxzdata}
The XXZ values in \eqref{eq:xxzdata} use the Pauli-matrix
normalization of \eqref{eq:K}, including its additive constant.
With \(\Delta=\cos\gamma=-1/2\) and \(\gamma=2\pi/3\),
the ground-energy integral \cite{yangyang} reduces to
\[
I=\frac3\pi\int_{-1}^{1}
\left(\frac1{1+z^2}-\frac2{3+z^2}\right)dz
=\frac32-\frac2{\sqrt3}.
\]
In this normalization, the energy density including the constant
in \eqref{eq:K} is
\[
\varepsilon_K=\cos\gamma+\frac16-2\sin\gamma\,I
=-\frac13-\sqrt3 I
=\frac53-\frac{3\sqrt3}{2}.
\]
The open-chain boundary term \cite{xxzboundary} cancels
between the two block minima as shown in
\eqref{eq:defect}.
\section{Periodic effective theory and the XXZ supercharge}\label{app:periodic}
The configuration-cycle calculation determines the positive
operator left by virtual spectator motion. We then match its
normalization to the periodic XXZ supercharge and derive the
spectral support of the correction. A comparison with open-chain
resolvents selects the periodic ground block and establishes the
low-energy and thermodynamic conclusions for the full effective theory.

\subsection{The kernel of a configuration cycle}
Put \(n=L-1\). For a cyclic active word \([w]\) of minimal period
\(p\mid n\), a spectator must make \(p\) circuits of the physical
ring to return to the same configuration. Its position graph is a
cycle with \(Lp\) vertices. In a sector with two or more spectators,
the zero-energy conditions from \eqref{eq:flag} can be propagated by
adjacent exchanges until two spectators touch; the \(00\) term then
sets that amplitude, and hence all amplitudes in the connected
sector, to zero. Thus only the all-active states and the alternating
one-spectator cycles contribute to the leading ground space.
For \(L\ge3\) in the uniform real gauge, its dimension is
\begin{equation}
g_L^{\rm per}=2^L+
\sum_{\substack{p\mid L-1\\ Lp\ {\rm even}}}M_2(p),\qquad
M_2(p)=\frac1p\sum_{d\mid p}\mu(d)2^{p/d}.
\label{eq:necklaces}
\end{equation}
Here \(\mu\) is the M\"obius function and \(M_2(p)\)
counts primitive binary necklaces of length \(p\).

Use \(\star\) for the spectator and relabel the active colors as
\(b=0,1\), with \(\bar b=1-b\). Each edge \(e\) joins two
configurations differing by \(\star b\leftrightarrow b\star\)
at a specified physical bond. Let \(\sigma_e\) be the all-active
word obtained by replacing that pair by \(\bar b\,\bar b\),
leaving all other sites fixed. On the orthonormal edge basis, define
\[
F_L^\dagger\ket e=\ket{\sigma_e},\qquad
F_L\ket\sigma=\sum_{e:\,\sigma_e=\sigma}\ket e.
\]
Different edges remain distinct even when their expanded words agree.
Let \(B\ket e\) be the sum of its two endpoint configurations.
Equation~\eqref{eq:collapse} gives
\[
H_1|_{\rm active}=BF_L,\qquad
F_L^\dagger F_L=\sum_{j=1}^L\Pi^{\rm equal}_{j,j+1}.
\]
The leading one-spectator operator is \(BB^\dagger\), and
\(BF_L\) is orthogonal to its kernel. The virtual resolvent
therefore acts on the positive-energy space.
For a cycle of length \(m\), \(Bx=0\) requires
\(x_{r+1}=-x_r\). The edge kernel is zero for odd \(m\), and for
even \(m\) is spanned by \((a_m)_r=(-1)^r\).
Singular-value decomposition gives
\[
B^\dagger(BB^\dagger)^+B=\I-\Pi_{\rm alt},\qquad
\Pi_{\rm alt}=
\bigoplus_{\substack{[w]\\Lp\ {\rm even}}}
\frac{\ket{a_{Lp}}\bra{a_{Lp}}}{Lp}.
\]
The leading resolvent preserves spectator number, whereas each
\(H_1\) changes it by one. Both the virtual term and \(H_0\)
therefore preserve its parity, so the all-active sector decouples
from the one-spectator leading ground states.
Subtracting the virtual term from the direct compression gives
\begin{equation}
H_{\rm eff,per}^{(0)}=K_L^{\rm per}+C_L,\qquad
C_L=F_L^\dagger\Pi_{\rm alt}F_L.
\label{eq:cyclecorrection}
\end{equation}
Since \(\Pi_{\rm alt}\) is an orthogonal projector, \(C_L\ge0\).

\subsection{Matching a normalized necklace to an insertion}
Let \(T_N\) be right translation and \(t_N=(-1)^{N+1}\).
The projector onto \(T_N=t_N\) is
\[
\Pi_N=\frac1N\sum_{j=0}^{N-1}(-1)^{(N+1)j}T_N^j.
\]
Fix the insertion signs and normalization by defining
\begin{align*}
\mathsf s\ket b&=\ket{1-b,1-b}/\sqrt2,\\
q_j^{(N)}&=(-1)^{j-1}\I^{\otimes(j-1)}
\otimes\mathsf s\otimes\I^{\otimes(N-j)},\qquad 1\le j\le N,\\
q_0^{(N)}&=-T_{N+1}^{-1}q_1^{(N)}T_N,\qquad
Q_N=\sqrt{\frac N{N+1}}\sum_{j=0}^Nq_j^{(N)}\Pi_N .
\end{align*}
These are the periodic dynamical-supercharge conventions of
Ref.~\cite{hagendorf}. In particular, the boundary term acts as
\[
q_0^{(N)}\ket{b_1\cdots b_N}
=-\frac1{\sqrt2}\ket{\bar b_N b_1\cdots b_{N-1}\bar b_N}.
\]
Translation sends \(q_j\) to \(-q_{j+1}T_N\) for
\(0\le j<N\), while \(T_{N+1}q_N=(-1)^Nq_0\).
On \(\Pi_N\), these relations give
\(T_{N+1}Q_N=-Q_NT_N\). Hence the image lies in \(\Pi_{N+1}\).

A necklace of minimal period \(p\) contributes to \(\Pi_n\)
precisely when \(t_n^p=1\), or \(Lp\) is even. Put
\(v_r=T_n^{-r}w\). Its normalized momentum state is
\[
\ket{w;t_n}=\frac1{\sqrt p}
\sum_{r=0}^{p-1}t_n^r\ket{v_r}.
\]
For \(v_r=b_1\cdots b_n\), label the edges by
\begin{align*}
e(r,0):&\quad
b_n b_1\cdots b_{n-1}\star
\longleftrightarrow\star b_1\cdots b_n,\\
e(r,j):&\quad
b_1\cdots b_{j-1}\star b_j\cdots b_n
\longleftrightarrow
b_1\cdots b_j\star b_{j+1}\cdots b_n,
\qquad 1\le j\le n.
\end{align*}
The ordered edges \(e(r,0),\ldots,e(r,n)\) form a path from
\(v_{r-1}\star\) to \(v_r\star\). The next edge is
\(e(r+1,0)\), with \(r\) understood modulo \(p\).
Each edge determines its physical bond, hence \(j\). Deleting the
spectator recovers \(v_r\) from either endpoint for \(j\ge1\),
and from the endpoint beginning in \(\star\) for \(j=0\).
Thus \((r,j)\), \(0\le r<p\), \(0\le j\le n\), is a
bijection with the \(Lp\) edges of the cycle.
For \(p<n\), the rotation stabilizer has size \(n/p\): using all
\(n\) rotations would repeat every \(v_r\) that many times.
The sum over \(p\) distinct words and its factor \(p^{-1/2}\)
already account for this stabilizer.

In this orientation the alternating edge vector is
\[
\ket{a_{Lp}}=\sum_{r=0}^{p-1}\sum_{j=0}^n
(-1)^{rL+j}\ket{e(r,j)}.
\]
The explicit boundary action and the interior insertions both give
\(q_j^{(n)}\ket{v_r}=(-1)^{j+1}\ket{\sigma_{e(r,j)}}/\sqrt2\).
Since \(t_n^r=(-1)^{rL}\), including the two normalization
factors yields
\begin{equation}
Q_n\ket{w;t_n}
=-\sqrt{\frac n{2Lp}}\,F_L^\dagger\ket{a_{Lp}}.
\label{eq:necklaceimage}
\end{equation}
Changing the cycle origin changes only the overall sign.
The allowed necklace momentum states are an orthonormal basis of
\(\Pi_n\). Summing their image projectors therefore yields
\[
Q_nQ_n^\dagger
=\sum_{\substack{[w]\\Lp\ {\rm even}}}
\frac n{2Lp}F_L^\dagger\ket{a_{Lp}}\bra{a_{Lp}}F_L
=\frac n2C_L.
\]
This proves the coefficient in \eqref{eq:supercharge}
and its stated momentum support.

\subsection{Nilpotence, commutation and a norm bound}
For either binary basis vector,
\begin{equation}
(\mathsf s\otimes\I-\I\otimes\mathsf s)\mathsf s\ket b
=\ket\chi\otimes\ket b-\ket b\otimes\ket\chi,\qquad
\ket\chi=-\frac12(\ket{01}+\ket{10}).
\label{eq:curvature}
\end{equation}
For \(b=0\), both sides equal
\((\ket{001}-\ket{100})/2\); the other case follows by color
exchange. Insertions at distinct positions cancel in the signed sum.
The remaining terms telescope by \eqref{eq:curvature}.
Including the two boundary terms, the unnormalized remainder on
\(\psi\) is
\[
T_{N+2}\bigl((T_N-t_N)\psi\otimes\chi\bigr)
+(t_NT_N-\I)\psi\otimes\chi.
\]
It vanishes on \(T_N\psi=t_N\psi\), proving
\(Q_{N+1}Q_N=0\).

The same cancellation between separated insertion and deletion
events leaves the two-site density
\begin{align}
h(\mathsf s)={}&
-(\mathsf s^\dagger\otimes\I)(\I\otimes\mathsf s)
-(\I\otimes\mathsf s^\dagger)(\mathsf s\otimes\I)
+\mathsf s\mathsf s^\dagger\nonumber\\
&+\frac12(\mathsf s^\dagger\mathsf s\otimes\I
+\I\otimes\mathsf s^\dagger\mathsf s).
\end{align}
On \(\Pi_L\), it satisfies
\[
\mathcal H_L^{\rm susy}:=Q_L^\dagger Q_L+Q_{L-1}Q_{L-1}^\dagger
=\Pi_L\sum_jh(\mathsf s)_{j,j+1}\Pi_L .
\]
Since \(\mathsf s^\dagger\mathsf s=\I/2\), direct multiplication gives
\[
h(\mathsf s)=
\begin{pmatrix}
1&0&0&0\\0&1/2&-1&0\\0&-1&1/2&0\\0&0&0&1
\end{pmatrix}
=\frac34\I-\frac12(XX+YY)+\frac14ZZ .
\]
Thus
\[
K_L^{\rm per}|_{\Pi_L}=-2\mathcal H_L^{\rm susy}+\frac{5L}{3}\I .
\]
Nilpotence makes the products, in either order, of the two positive
terms in \(\mathcal H_L^{\rm susy}\) vanish. In particular,
\(Q_{L-1}Q_{L-1}^\dagger\) commutes with it. On the other momentum
sectors \(C_L=0\), proving the full commutation statement.

The local eigenvalues of \(h(\mathsf s)\) are
\(1,1,3/2,-1/2\). Positivity of the supercharge squares gives
\[
0\le Q_{L-1}Q_{L-1}^\dagger\le\mathcal H_L^{\rm susy},
\qquad
\|C_L\|\le\frac2{L-1}\|\mathcal H_L^{\rm susy}\|
\le\frac{3L}{L-1}.
\]
The min--max principle bounds each ordered eigenvalue shift by this
quantity. For any fixed \(\beta>0\), the partition functions satisfy
\[
e^{-3\beta L/(L-1)}Z(K_L^{\rm per})
\le Z(K_L^{\rm per}+C_L)\le Z(K_L^{\rm per}).
\]
Taking logarithms and dividing by \(\beta L\) proves equality of
the thermodynamic free-energy densities of the two effective operators.

\subsection{Spectral support and a protected energy interval}
Let \(e_m^{\rm per}=\min\spec K_m^{\rm per}\).
Nilpotence also gives the intertwining relation
\(\mathcal H_L^{\rm susy}Q_{L-1}
=Q_{L-1}\mathcal H_{L-1}^{\rm susy}\). Hence
\begin{equation}
K_L^{\rm per}Q_{L-1}
=Q_{L-1}\left(K_{L-1}^{\rm per}+\frac53\I\right).
\label{eq:adjacentlength}
\end{equation}
The nonzero support of \(C_L\) is \(\operatorname{im}Q_{L-1}\).
On this image, \(\mathcal H_L^{\rm susy}=Q_{L-1}Q_{L-1}^\dagger\).
Thus a \(K_L^{\rm per}\) eigenvector of energy \(E\) in this image
has correction eigenvalue
\begin{equation}
c_L(E)=\frac{5L/3-E}{L-1},\qquad
E\ge e_{L-1}^{\rm per}+\frac53.
\label{eq:spectralsupport}
\end{equation}
The inequality follows by applying \(Q_{L-1}^\dagger\) to the
eigenvector in \eqref{eq:adjacentlength}; its image is nonzero.

Put \(n=L-1\). Average an \(n\)-site XXZ ground-state density
matrix over translations and insert a maximally mixed spin into
one bond. Removing that bond costs its mean energy
\(e_n^{\rm per}/n\), and the two new bonds each have expectation
\(1/6\). The variational principle gives
\begin{equation}
e_L^{\rm per}\le\frac{n-1}{n}e_n^{\rm per}+\frac13.
\label{eq:periodicinsert}
\end{equation}
This remains valid for \(n=2\), where the periodic sum contains
two copies of the bond density.
For \(n\ge3\), the two adjacent bond densities have eigenvalues
\[
-\frac23,\quad\frac13,\quad
\frac{5-3\sqrt{33}}6,\quad\frac{5+3\sqrt{33}}6,
\]
each twice. Summing this three-site lower bound around the ring
gives
\(e_n^{\rm per}/n\ge(5-3\sqrt{33})/12\). Consequently,
\begin{equation}
e_{L-1}^{\rm per}+\frac53-e_L^{\rm per}
\ge\frac43+\frac{e_n^{\rm per}}n
\ge\delta_*,\qquad
\delta_*=\frac{7-\sqrt{33}}4>0,\qquad L\ge4.
\label{eq:protectedgap}
\end{equation}
Equations~\eqref{eq:spectralsupport} and \eqref{eq:protectedgap}
show that \(C_L\) vanishes on every XXZ eigenstate with energy
below \(e_L^{\rm per}+\delta_*\).
Since \(C_L\) is positive and commutes with \(K_L^{\rm per}\),
the two operators have identical spectra, with multiplicities,
throughout that interval. In particular their ground energies
and ground-state spaces agree for \(L\ge4\).
At \(L=3\), direct evaluation gives
\[
\spec K_3^{\rm per}=\{-1\ (6),\ 5\ (2)\},\qquad
C_3=3\ket{\psi_-}\bra{\psi_-},\qquad
\ket{\psi_-}=\frac{\ket{111}-\ket{222}}{\sqrt2},
\]
where parentheses denote multiplicities. The correction raises
\(\ket{\psi_-}\) from energy \(-1\) to \(2\).
The common ground energy \(-1\) therefore has degeneracy six
for XXZ and five for the all-active effective block.

\subsection{A lower bound on the periodic spectator block}
The one-spectator ground space has an isometric representation
on \(\Pi_n\mathcal H_n\), with \(n=L-1\). Define
\[
W_L\ket w=\frac1{\sqrt L}\sum_{p=0}^{n}(-1)^p\ket{p;w},
\]
where \(\ket{p;w}\) inserts \(\star\) after the first \(p\)
letters of the active word \(w\). Then \(W_L^\dagger W_L=\I\).
The alternating signs enforce every internal leading bond
condition; the closing bond imposes \(T_n=t_n\).
Thus \(W_L\Pi_n\) maps onto the full one-spectator ground space.
If \(\mathsf T_L\) denotes physical qutrit translation, then
\[
\mathsf T_LW_L\Pi_n=(-1)^{L+1}W_L\Pi_n.
\]
Indeed an internal spectator position acquires one minus sign
and an active-word rotation, whereas a spectator at the last
site contributes \((-1)^n\); the two factors agree on \(\Pi_n\).

Delete bond \(c\) of the ring and denote the resulting open-chain
coefficients by \(H_{k,c}^{\rm op}\), for \(k=0,1,2\).
The periodic ground space is contained in each open ground space.
On the two-spectator space, set
\[
A_c=H_{2,c}^{\rm op}+\frac{2n}{3}\I,\qquad
A=H_2^{\rm per}+\frac{2L}{3}\I,
\]
and let \(S_c\) and \(S\) be the two-spectator components of
\(H_{1,c}^{\rm op}W_L\Pi_n\) and \(H_1^{\rm per}W_L\Pi_n\).
Their all-active components vanish by the alternating cancellation
in Appendix~\ref{app:open}. Each \(A_c\) is positive definite by
\eqref{eq:exterior}. Every bond occurs in \(n\) of the cuts, so
\[
\sum_{c=1}^{L}A_c=nA,\qquad
\sum_{c=1}^{L}S_c=nS.
\]
For positive \(A_c\), the matrix Cauchy--Schwarz inequality gives
\begin{equation}
S^\dagger A^{-1}S
\le\frac1n\sum_{c=1}^{L}S_c^\dagger A_c^{-1}S_c.
\label{eq:cutresolvent}
\end{equation}
To derive this inequality, expand
\(\sum_c\|A_c^{-1/2}S_c\psi-A_c^{1/2}z\|^2\ge0\)
with \(z=(\sum_cA_c)^{-1}\sum_cS_c\psi\).

The direct term is also \(1/n\) times the sum over cuts.
Moreover, translation acts as a scalar on \(W_L\Pi_n\), so all
compressed open effective operators in this sum coincide.
Using \eqref{eq:blocks} yields, on \(\Pi_n\mathcal H_n\),
\begin{align}
H_{\rm eff,per}^{(1)}
&\ge\frac{L}{n}\Pi_n
\left(\frac{10n}{3L}\I+\frac nL K_n\right)\Pi_n\nonumber\\
&=\frac{10}{3}\I+
\frac{n-1}{n}\left.K_n^{\rm per}\right|_{\Pi_n}.
\label{eq:periodicspectatorbound}
\end{align}
The final step uses translation within \(\Pi_n\): the compression
of the \(n-1\) open XXZ bonds is \((n-1)/n\) times the periodic sum.
Taking minimum energies and applying \eqref{eq:periodicinsert} gives
\begin{equation}
\min\spec H_{\rm eff,per}^{(1)}-e_L^{\rm per}
\ge3,\qquad L\ge3.
\label{eq:periodicspectatorgap}
\end{equation}
Since the all-active block has ground energy \(e_L^{\rm per}\),
this selects the ground block of the complete periodic effective
Hamiltonian. Together with \eqref{eq:protectedgap}, it proves the
full low-energy spectral equality \eqref{eq:fullperiodiclowenergy}.

The same bound controls the additional sector at positive
temperature. If \(Z_m^K(\beta)=\Tr e^{-\beta K_m^{\rm per}}\),
the min--max principle and \eqref{eq:periodicspectatorbound} give
\[
Z_L^{(1)}(\beta)\le e^{-10\beta/3}
Z_n^K\left(\beta\frac{n-1}{n}\right).
\]
The temperature rescaling changes \(\log Z_n^K\) by \(O(1)\),
since \(\|K_n^{\rm per}\|=O(n)\).
The all-active partition function has the XXZ pressure by
\eqref{eq:Cbound}; adding \(Z_L^{(1)}\) leaves its limiting
pressure unchanged. The full periodic effective Hamiltonian
therefore has the same bulk free-energy density as XXZ.
\section*{CRediT authorship contribution statement}
\textbf{Hanbing Liang:} Conceptualization, Project administration, Validation,
Writing -- original draft.
\textbf{Fujun Liu:} Validation, Writing -- review \& editing.

\section*{Funding}
This work was supported by the Science and Technology Development Project
of Jilin Province (Grant No.\ 20250102032JC).

\section*{Declaration of competing interest}
The authors declare that they have no competing interests.

\section*{Data availability}
No data was used for the research described in the article.

\section*{Code availability}
The arXiv ancillary files contain \texttt{verify\_symbolic.py},
\texttt{verify\_transport.py}, their dependencies and usage instructions.
The scripts use Python and SymPy for the exact checks described in
Appendices~\ref{app:spectral} and \ref{app:transport}.
Optional checks compare the open-chain effective blocks at \(L=2,3,4\)
and the periodic all-active block at \(L=3,4\) with the analytical
expressions in Appendices~\ref{app:open} and \ref{app:periodic}.
They also check the adjacent-length supercharge relation and the
periodic spectator operator bound at \(L=3,4\), and the local
three-site spectrum used to prove low-energy protection.
The LaTeX source includes the PGFPlots and TikZ code for both figures.

\section*{Declaration of generative AI and AI-assisted technologies
in the manuscript preparation process}
OpenAI Codex (\mbox{OpenAI}), using models \texttt{gpt-6-astra} and
\texttt{gpt-5.6-sol}, assisted with manuscript preparation and language
editing. Model \texttt{gpt-6-astra} also assisted with developing the
symbolic verification scripts and the LaTeX source, including the
PGFPlots and TikZ figure code. Research applications are described in
Appendix~\ref{app:methods}.
The authors have checked and revised the AI-assisted material,
verified the final manuscript, and accept full responsibility for its content.

\end{document}